\documentstyle[11pt]{article}
\def\beq{\begin{equation}}
\def\eeq{\end{equation}}
\def\beqa{\begin{eqnarray}}
\def\eeqa{\end{eqnarray}}

\title{The Nuclear Born Oppenheimer Method and Nuclear Rotations}
\author{Nouredine Zettili}
\date{Department of Physical \& Earth Sciences\\ Jacksonville State
University\\ Jacksonville, AL  36265}

\begin{document}
\sf

\maketitle
\begin{abstract}
 We deal here with the application of the Nuclear Born Oppenheimer (NBO)  method   to the description of nuclear rotations. As an edifying illustration, we apply the NBO formalism to study the rotational motion of nuclei which are axially-symmetric and  even, but whose shells are  not  closed. We  focus, in particular, on the derivation of expressions for the rotational energy and for the moment of inertia. Additionally, we  examine the connection between the NBO method and the  self-consistent cranking  (SCC)  model.  Finally, we compare  the moment of inertia generated by the NBO method  with  the Thouless-Valantin formula and hence establish a connection between the NBO method and the large body of experimental data.

\vspace*{0.3cm}

\noindent PACS numbers: 21.60.-n, 21.60.Ev, 21.10.Re

\end{abstract}
\newpage

\vspace*{10pt}

\section{Introduction}
\vspace*{-0.5pt}

\noindent Since nuclear and molecular rotation-vibration spectra present many striking analogies, and since the Born-Oppenheimer (BO) approximation\cite{bo1} of molecular physics was shown to be very accurate\footnote{Using an elementary
solvable model, Moshinksy and Kittel\cite{mos1} have shown that
the BO approximation is very accurate for both the molecular
energy and wave function: ${E_{\rm BO}\over E_{\rm exact}} = 1 -
{1\over 4} \chi$ and $|\langle\psi_{\rm BO}|\psi_{\rm
exact}\rangle|^2 = 1-{3\over 128} \chi^2$, where $\chi$ is equal
to the ratio of electronic to nuclear masses (i.e., $\chi=m_e/M\simeq 10^{_4}$).} in describing molecular rotations and vibrations\cite{mos1}, it will be interesting to explore the possibility of using the BO approximation to describe nuclear collective rotations.

Exploiting the analogy between nuclear and molecular dynamics,
Villars introduced a microscopic method\cite{vil1}, \cite{vil2} to
describe nuclear collective motion. This method, to be called the
Nuclear Born-Oppenheimer (NBO) method, was developed along the
lines of the molecular BO approximation by constructing a
factorable trial function modeled after the  BO ansatz.

Using an analytically solvable model\cite{zet1}, we have shown
that the NBO method is very accurate for adiabatic collective
motion\cite{zet2}. Since the NBO method is a quantum mechanical
prescription, we have shown that it offers a suitable framework
for describing the zero-point fluctuations\cite{zet3}; we have
also shown that  the method offers an accurate description of
small-amplitude collective oscillations\cite{zet4} and that it
yields the random phase approximation (RPA) equations\cite{zet5}. Additionally, we  have applied  the NBO method to study nuclear
collective motion\cite{zet6} and examined its connection with the collective
model of Bohr\cite{bohr}.

So, having applied the NBO method to the study of small amplitude motion, we have yet to apply it to nuclear collective rotations. In this work we want to achieve just that aim; namely, we want to apply the general BO formalism outlined in Ref.\cite{zet6}  to the description of nuclear rotational states. As an illustration, we will apply the NBO formalism to study the rotations of nuclei that are axially-symmetric and even,  but with non-closed shells. We will focus, in particular, on the derivation of
expressions for the energy and for the moment of inertia. Additionally, we
shall examine the connection of the NBO method with the successful
self-consistent cranking (SCC) model.

In Sec. 2, we present a brief outline of the  NBO formalism
and   how it applies  to nuclear nuclear collective    motion.
 We then devote Sec. 3 to the application of the NBO method to study the rotational states of axially symmetric nuclei; in particular, we will derive an expression for the rotational  energy.   In Sec. 4, we present a discussion on the connection of the NBO method to the self-consistent cranking (SCC) model.


\vspace*{1pt}
\section{Synopsis of the Application of the NBO  Method to Nuclear Collective Motion}
\vspace*{-0.5pt} \noindent To describe nuclear collective motion within
the framework of the NBO method,   we need to introduce a tensor
operator $\hat Q_{\alpha\beta}$; that is, to be able to describe collective
rotations and vibrations of nuclei, we need to introduce a set of
operators that are the elements of a symmetric cartesian tensor
operator $\hat Q_{\alpha\beta}$. In the rest of this work, we
shall use   Greek subscripts  to refer to a  space-fixed frame of
reference; Latin subscripts  will be used later to refer  to
a body-fixed frame. The  operator $\hat Q_{\alpha\beta}$  is
assumed to depend on the various nucleonic variables -- positions $\vec x_i$,
momenta, $\vec p_i$, and spins, $\vec s_i$.  In addition, we
assume that $\hat Q_{\alpha\beta}$ are one-body operators,
symmetric, even under time reversal, have a continuous eigenvalue
spectrum $(q_{\alpha\beta})$, and commute  with any other
component $\hat Q_{\gamma\delta}$ of $\hat Q$ (\emph{ i.e.},
$[\hat Q_{\alpha\beta},\ \hat Q_{\gamma\delta}] = 0$). Let $\hat
K_{\alpha\beta}$ be the canonical  conjugate of $\hat
Q_{\alpha\beta}$:
\begin{equation}
[ i\hat K_{\alpha\beta},\ \hat Q_{\gamma\delta}] = {1\over 2}
\left( \delta_{\alpha\gamma} \delta_{\beta\delta} +
\delta_{\alpha\delta}\delta_{\beta\gamma}\right)
.\label{eq:c7.2-1}
\end{equation}

The NBO method  consists of the following two essential
steps\cite{vil1}, \cite{vil2}, \cite{zet6}:
\begin{itemize}
\item First, we need to construct a suitable representation for the nucleus' Hamiltonian $\hat H$ by   decomposing it  into a series
\begin{equation}
\hat H = \hat H_0 + \sum_{\alpha\beta} \hat H_{1\,\alpha\beta}
\hat K_{\alpha\beta} + {1\over 2}
\sum_{\alpha\beta}\sum_{\gamma\delta} \hat
H_{2\,\alpha\beta\gamma\delta} \hat K_{\alpha\beta}\hat
K_{\gamma\delta} + \cdots, \label{eq:c7.2-2}
\end{equation}
where all coefficient operators $\hat H_0$, $\hat
H_{1\,\alpha\beta}$,  $\hat H_{2\,\alpha\beta,\gamma\delta},\ldots$
commute with $\hat Q_{\alpha\beta}$.

\item   Second, we make use a factorable  trial function\footnote{We will use $x$ to abbreviate
for the set of nucleonic variables -- position $\vec x_i$, momentum $\vec p_i$, and spin, $\vec s_i$.}
\begin{equation}
\langle x|\psi\rangle = \int \prod_{\alpha\le\beta}
dq_{\alpha\beta} \langle x|\delta(q_{\alpha\beta} - \hat
Q_{\alpha\beta}) |\Phi (q) \rangle g(q)\ \ ,\label{eq:c7.2-3}
\end{equation}
where $\langle x|\Phi(q)\rangle$ is the intrinsic wave function,
and $g(q)$ is the collective amplitude.

\end{itemize}

After constructing the Hamiltonian and the  wave function, we can calculate the mean energy by a simple application of $\hat H$ to $|\psi\rangle$:
\begin{eqnarray}
\langle\psi|\hat{H}|\psi\rangle \!\!\!\!\! & = & \!\!\!\!\!  \int
\prod_{\alpha\le\beta} dq_{\alpha\beta}\
g^*(q)\langle\Phi(q)|\delta(q_{\alpha\beta} - \hat
Q_{\alpha\beta}) \bigl\{ \tilde H_0|\Phi\rangle g(q)  +
\sum_{\alpha\beta} \tilde H_{1\,\alpha\beta} |\Phi\rangle
k_{\alpha\beta} g(q) \nonumber \\
& &\quad + {1\over 2}\sum_{\alpha\beta}\sum_{\gamma\delta} \tilde
H_{2\,\alpha\beta,\gamma\delta} |\Phi\rangle k_{\alpha\beta}
k_{\gamma\delta} g(q) + \cdots\bigr\}  , \label{eq:c7.2-4}
\end{eqnarray}
where the $k_{\alpha\beta}$ are operators that act on $g(q)$; they obey commutation
relations with the $q_{\alpha\beta}$ isomorphic with (\ref{eq:c7.2-1})
\begin{equation}
\left[ ik_{\alpha\beta},\ q_{\gamma\delta}\right] = {1\over 2}
\left( \delta_{\alpha\gamma} \delta_{\beta\delta} +
\delta_{\alpha\delta} \delta_{\beta\gamma}\right)
.\label{eq:c7.2-5}
\end{equation}
The few lowest expressions of  $\tilde H_K$ are given by
\begin{eqnarray}
\!\!\!\!\!\!\!\!\!\! \tilde H_0 \!\!\!\!\! & = & \!\!\!\!\! \hat H
- \sum_{\alpha\beta} \left[ \hat H,\ i\hat Q_{\alpha\beta}\right]
\hat G_{\alpha\beta} + {1\over 2}
\sum_{\alpha\beta}\sum_{\gamma\delta} \left[\left[ \hat H,\  i
\hat Q_{\alpha\beta}\right],\  i \hat Q_{\gamma\delta}\right] \hat
G_{\alpha\beta} \hat
G_{\gamma\delta} + \cdots ,\nonumber \\
\tilde H_{1\,\alpha\beta} \!\!\!\!\! & = & \!\!\!\!\!    \left[
\hat H,\  i \hat Q_{\alpha\beta}\right] -\sum_{\gamma\delta}
\left[\left[ \hat H,\ i\hat Q_{\alpha\beta}\right],\ i\hat
Q_{\gamma\delta}\right]
\hat G_{\gamma\delta}+ \cdots   , \qquad \label{eq:c7.2-6}   \\
\tilde H_{2\,\alpha\beta\;\gamma\delta} \!\!\!\!\! & = &
\!\!\!\!\!   \left[\left[ \hat H,\ i\hat Q_{\alpha\beta}\right],\
i \hat Q_{\gamma\delta}\right] + \cdots ,\nonumber
\end{eqnarray}
where $\hat G_{\alpha\beta}$ is a one particle operator that acts on $|\phi\rangle$; it is defined
by the action of $k_{\alpha\beta}$ on the parameter $q$ in $|\phi\rangle$
\begin{equation}
k_{\alpha\beta} \langle x|\Phi(q)\rangle = {1\over 2i} (1 +
\delta_{\alpha\beta}) {\partial\over\partial q_{\alpha\beta}}
\langle x|\Phi(q) \rangle = \langle x|\hat
G_{\alpha\beta}|\Phi(q)\rangle . \label{eq:c7.2-7}
\end{equation}

We should note that the  mean energy expression (\ref{eq:c7.2-4}) was derived
within a  space-fixed or lab frame.  However, in the description of permanently
deformed (non spherical) nuclei,  it is more convenient to employ a
body-fixed  frame of reference. Here, we take the axes of the
body-fixed frame along the three principal axes of
$q_{\alpha\beta}$ which are defined by the unit vectors $\hat e_a$
$(a=1,2,3)$, and specify their orientation with respect to the
space-fixed frame by three Euler angles\cite{zet10} $\theta_s$ (i.e.,
$\theta,\varphi,\psi)$: $\hat e_a=\hat e_a(\theta_s$) with $\hat
e_a\cdot\hat e_b=\delta_{ab}$ and $\hat e_a\times\hat e_b={\cal
E}_{abc}\hat e_c$ where ${\cal E}_{abc}$ is the antisymmetric
tensor  (${\cal E}_{123}=1=-{\cal E}_{213}$ etc.). The collective
degrees of freedom can be separated into rotational and
vibrational terms by transforming $\hat Q_{\alpha\beta}$   to the
body-fixed frame; that is, by means of the principal axes
transformation of the tensor operator $\hat Q_{\alpha\beta}$:
\begin{equation}
q_{\alpha\beta} = \sum^3_{a=1} e_{\alpha a}(\theta) e_{\beta
a}(\theta) q_a
 ,\label{eq:c7.2-8}
\end{equation}
where $e_{\alpha a} (\equiv \hat e_\alpha \cdot \hat e_a)$, the
$\alpha^{\rm th}$ component of the unit vector $\hat e_a$, depends
on the three Euler angles $\theta_s$.  In the transformation to
the  body-fixed frame, we have essentially replaced the six
collective coordinates $q_{\alpha\beta}$ by the three $q_a$'s and
the three Euler angles. The matrices $e_{\alpha\beta}$ obey the
orthogonality relations: $\sum_a e_{\alpha a} e_{\beta a} =
\delta_{\alpha\beta}$ and $\sum_\alpha e_{\alpha a} e_{\alpha b} =
\delta_{ab} $.

We can now introduce rotation
operators ${\hat{\cal L}}_a$ about
the body-fixed axes. The Euler angles specifying the orientation
of the intrinsic frame need to be viewed as dynamical variables;
for instance, the unit vector $\hat e_a$ satisfy the commutation
rules of a vector operator
\begin{equation}
\left[ {\hat{\cal L}}_{[ab]},\ e_{\alpha c}\right] = ie_{\alpha a}
\delta_{bc} - ie_{\alpha b} \delta_{ac} . \label{eq:c7.2-10}
\end{equation}
We  can easily verify from (\ref{eq:c7.2-10}) that these operators
obey the commutation relations
\begin{equation}
\left[ {\hat{\cal L}}_a,\  {\hat{\cal L}}_b\right] = - i{\cal E}_{abc}{\hat{\cal L}}_c ,\label{eq:c7.2-11}
\end{equation}
which differ in sign from the commutation rules
of ordinary angular momentum\cite{blrpk} because they refer to the
moving axes and hence do not have the same commutation properties
as angular-momentum components along space fixed axes. For instance,
we have  $\left[{\hat{\cal L}}_1,\ {\hat{\cal L}}_2\right] =-i{\hat{\cal L}}_3$. The space fixed components ${\hat{\cal L}}_{\alpha\beta}$ of $\vec{\cal
L}$ can be obtained by rotation: ${\hat{\cal L}}_{\alpha\beta} =
\sum_{ab} e_{\alpha a} e_{\beta b} {\hat{\cal L}}_{ab}$.

In conjunction with the replacement of $q_{\alpha\beta}$ by the variables
$q_a$ and $\theta_s$, we seek an expression for the operator
$k_{\alpha\beta}$ in terms of the ${\hat{\cal L}}_{\alpha\beta}$ and a set of three
operators $p_a$ conjugate to $q_a$, with $\left[ p_a,\ q_b\right] = i\delta_{ab}$.
We can verify\cite{zet6} that  $k_{\alpha\beta}$ transforms like an operator that acts
on $\theta_s$ and $q_a$:
\begin{eqnarray}
k_{\alpha\beta} & = &  {1\over 2} \sum_{ab} {e_{\alpha a} e_{\beta b}
\over q_a - q_b} {\hat{\cal L}}_{[ab]} + \sum_a e_{\alpha a} e_{\beta a}
p_a \nonumber \\ & = &   \sum_{ab} e_{\alpha a} e_{\beta b} \left[ {1\over 2} \left(
1+ \delta_{\alpha\beta}\right) {{\hat{\cal L}}_{[ab]}\over q_a-q_b} +
\delta_{ab} p_a\right] . \label{eq:c7.2-17}
\end{eqnarray}
Additionally, we can ascertain that $k_{\alpha\beta}$ is Hermitian with regard to the volume 
element $\prod_{\alpha\le\beta} dq_{\alpha\beta}$, which can be shown to transform 
like:
\begin{equation}
\int d\tau=\int \prod_{\alpha\le\beta} dq_{\alpha\beta} = \int
\prod^3_{a=1} dq_a (q_1-q_2) (q_2-q_3)(q_3-q_1) d\Omega
,\label{eq:c7.2-18}
\end{equation}
where $d\Omega$ is the usual angular element\cite{zet10} $d\Omega= \sin\theta
d\theta d\varphi d\psi$.

Using the relations (\ref{eq:c7.2-17}) and (\ref{eq:c7.2-18}), we
can now express (\ref{eq:c7.2-4}) and (\ref{eq:c7.2-6}) in the body-fixed frame.
For this, note first that under the
transformation (\ref{eq:c7.2-8}) from the Lab frame to the
body-fixed system, the quantities $g(q_{\alpha\beta})$, $\langle
x_{i\alpha} s_{i\alpha}|\Phi(q_{\alpha\beta})\rangle$, and
$\prod_{\alpha\le\beta}\delta(q_{\alpha\beta} - \hat
Q_{\alpha\beta})$ become $f(q_a,\theta_s)$, $\langle x'_{ia}
s'_{ia}|\phi(q_a)\rangle$, and $\prod_a\delta(q_a - \hat
Q_{aa})\prod_{a\le b} \delta (Q_{ab})$, respectively, which in
turn will be abbreviated to $f(q,\theta)$, $\langle x'_{i}
s'_{i}|\phi(q)\rangle$, and $\delta(q - \hat Q)$. 

Next, we can show\cite{zet6} that the action of the total
angular momentum $\hat{J}_{\alpha\beta}$ on $|\psi\rangle$ can be
expressed in terms of ${\hat{\cal L}}_{\alpha\beta}$ on the collective
amplitude $f(q_a,\theta_s)$:
\begin{equation}
\langle x|\hat J_{\alpha\beta}|\psi\rangle = \int d\tau\langle
x|\delta(q - \hat Q)|\phi(q)\rangle {\hat{\cal L}}_{\alpha\beta}
f(q,\theta) . \label{eq:c7.2-26}
\end{equation}
In this new representation, the operator $\hat G_{\alpha\beta}$ of
(\ref{eq:c7.2-6}) is
rotated into $\hat G_{ab}$:
\begin{eqnarray}
\langle x'_{ia} s'_{ia}|\hat G_{ab}|\phi(q_a)\rangle = {1\over 2} (1-\delta_{ab})
{\langle x'_{ia}s'_{ia}|\hat J_{[ab]}|\phi(q_a)\rangle \over q_a - q_b} + \delta_{ab}
\langle x'_{ia}s'_{ia}|\hat G_a|\phi(q_a)\rangle .\nonumber \\
\label{eq:c7.2-27}
\end{eqnarray}

Finally,   using Eq.~(\ref{eq:c7.2-17}), (\ref{eq:c7.2-26}) and
(\ref{eq:c7.2-27}),  we have shown in Ref.\cite{zet6} that the mean energy (\ref{eq:c7.2-4}) is given in the body-fixed frame of reference by\footnote{\, Recall that $\hat J_{[ab]}$ and $\hat G_a$ operate
on the intrinsic state $|\phi(q_a)\rangle$, but ${\hat{\cal L}}_{[ab]}$
and $p_a$ operate on the collective state $f(q_a,\theta_s)$ (i.e.,
${\hat{\cal L}}_{[ab]}$ acts on $\theta_s$ and $p_a$ on $q_a$).}
\begin{eqnarray}
\langle\psi|\hat H|\psi\rangle  \!\!\!\! & = & \!\!\!\! \int d\tau
f^*(q,\theta) \langle\phi(q)|\delta (q-\hat Q) \biggl\{ \hat H -
\sum^3_{a,b=1} \dot{\hat Y}_{[ab]} (\hat J_{[ab]} - {\hat{\cal L}}_{[ab]})\nonumber \\
& &    -\ \sum^3_{a=1} \dot{\hat Q}_a (\hat G_a-p_a)   +   {1\over
2} \sum_{abcd} \hat B_{[ab],[cd]} (\hat J_{[ab]} - {\hat{\cal L}}_{[ab]}) (\hat J_{[cd]} - {\hat{\cal L}}_{[cd]})  \nonumber \\
& & +\ \sum_{abc} \hat B_{[ab],c} (\hat J_{[ab]} - {\hat{\cal L}}_{[ab]}) (\hat G_c - p_c)
\nonumber \\
& & +\  {1\over 2} \sum_{ab} \hat B_{a,b} (\hat G_a - p_a)(\hat
G_b - p_b) \biggr\}|\phi(q) \rangle f(q,\theta) ,
\label{eq:c7.2-28}
\end{eqnarray}
where
\begin{equation}
\hat Y_{[ab]} = {\hat Q_{ab} \over q_a-q_b},\qquad \dot{\hat
Y}_{[ab]} = \left[ i\hat H,\ \hat Y_{[ab]}\right]\ ,\quad
\dot{\hat Q}_a = \left[ i\hat H,\ \hat Q_a\right]\
,\label{eq:c7.2-29}
\end{equation}
\begin{equation}
\hat B_{[ab],[cd]} =  [ \dot{\hat Y}_{[ab]},\ i\hat Y_{[cd]} ]\ ,
\quad \hat B_{[ab],c} =  [  \dot{\hat Y}_{[ab]},\ i\hat Q_c ]\ ,
\quad \hat B_{ab} =  [ \dot{\hat Q}_a,\ i\hat Q_b ] .
\label{eq:c7.2-30}
\end{equation}
Note that,  in deriving the mean energy
(\ref{eq:c7.2-28}), we have terminated the series
(\ref{eq:c7.2-4}) at the quadratic terms in $k_{\alpha\beta}$.  This
termination is justified by the validity of the adiabatic
approximation in the present case, since we are dealing with
nuclear dynamics for which the time evolution of the collective
variables is assumed to be slow on the scale of a single-particle
(nucleonic) motion.

As we are going to see next,   the rotational and
vibrational  degrees of freedom appear explicitly in the energy
expression (\ref{eq:c7.2-28}); we will also show how to derive expressions for
the collective rotational energy and for the moment of inertia.

\section{Description of Rotational States of Axially-Symmetric Nuclei}

Consider a permanently deformed, non spherical nucleus. Since we are
interested in   rotational motion only, we  assume the
nucleus to be in its vibrational ground state. In this case, we assume that
the collective amplitude $f(q_a,\Omega)$ of (\ref{eq:c7.2-28}) separates into a
vibrational part, $g_0(q_a-\bar q_a)$, and a rotational part, $D(\Omega)$:
\begin{equation}
f(q_a,\Omega) = g_0(q_a-\bar q_a)D(\Omega).
\end{equation}
The vibrational collective amplitude $g_0(q_a-\bar q_a)$ represents here the
zero-point oscillations about the equilibrium values, $\bar q_a$, of $q_a$ $%
(a=1,2,3)$. Hence, the wave function $|\psi\rangle$ of the system becomes
(\textit{c.f.} Eq. (\ref{eq:c7.2-3})):
\begin{equation}
|\psi\rangle = \int d\tau d\Omega\ \delta(q-\hat Q)|\phi(q_a)\rangle
g_0(q_a-\bar q_a) D(\Omega)\ \ ,  \label{eq:9.2-1}
\end{equation}
with $d\tau = (q_a - q_2)(q_2-q_3)(q_3-q_1)\prod^3_{a=1}dq_a$ and $d\Omega =
d\varphi d\psi\sin\theta d\theta$ (\textit{c.f.} Eq. (\ref{eq:c7.2-18})), and where $%
\delta(q- \hat Q)$ is used to abbreviate $\prod^3_{a=1}\delta(q_a - \hat
Q_{aa})\prod_{a<b} \delta(\hat Q_{ab})$. In this case, after expanding $%
|\phi(q_a)\rangle$ about $|\phi(\bar q_a)\rangle$, the mean energy (\ref{eq:c7.2-28})
becomes:
\begin{eqnarray}
\langle\psi|\hat{H}|\psi \rangle \!\!\!\! & \simeq &\!\!\!\!\! \int d\tau\int
d\Omega g^*(q_a - \bar q_a) D^*(\Omega) \langle\phi(\bar q_a)|\delta(q- \hat Q) %
\biggl\{ \hat H - \sum^3_{a=1} \dot {\hat Y}_a(\hat J_a - {\hat{\mathcal{L}}}_a) %
\biggr.  \nonumber \\
& & \biggl. +\ \ {\frac{1}{2}} \sum_{ab} B_{ab} (\hat J_a - {\hat{\mathcal{L}}}_a)
(\hat J_b - {\hat{\mathcal{L}}}_b) \biggr\} |\phi(\bar q_a)\rangle g(q_a - \bar q_a)
D(\Omega)  \nonumber \\
& &+ \ \ E^0_{\mathrm{osc}} + E_{\mathrm{coupl}},\ \   \label{eq:9.2-2}
\end{eqnarray}
where we have used the notation $\hat 0_c$ to abbreviate $\hat 0_{ab}$ ($%
a,b,c$ being cyclic permutations of the body-fixed axes 1,2,3). In this expression, $E^0_{%
\mathrm{osc}}$ is the energy of the zero-point oscillations; $E_{\mathrm{%
coupl}}$ is the coupling-energy between the rotational and vibrational
motions which we can neglect. As for $E^0_{\mathrm{osc}}$, we will drop it
from all following mean-energy expressions, since it represents only a
constant shift of the entire (rotational) energy spectrum. Note that, in the
derivation of the mean-energy (\ref{eq:9.2-2}), we have approximated  the operator $\hat B_{ab}$ by its mean value $\langle\phi|\hat
B_{ab}|\phi\rangle$ (i.e., $\langle\phi|\hat
B_{ab}|\phi\rangle\equiv B_{ab})$. In what follows, the notation $B_a$
will be used to abbreviate $B_{aa}$

For the sake of simplicity, we shall \textit{focus in this  work  only on
deformed,  even, and axially-symmetry nuclei.} Consider the axis 3, of the
body-fixed frame, to be the axis of symmetry for the system. As a
consequence of the axial symmetry, we have: $\bar q_a = \bar q_2\not= \bar
q_3$ and $B_1 = B_2 \equiv B\not= B_3$.

Now, since $\hat{H}$, ${\hat{\vec J}}\,^2$ and $\hat{J_z}$ mutually commute\footnote{$\vec J$ is the total angular
momentum and $\hat{J_z}$ is its $Z$ component with respect to the Lab frame.}
commute, they possess joint eigenfunctions. The structure of our trial
function allows it to be an exact eigenfunction of ${\hat{\vec J}}\,^2$ and $\hat{J_2}$, but
provides only a variational approximation to the energy. In the case of
axial symmetry, this trial function $|\psi\rangle$ can be obtained from (\ref%
{eq:9.2-1}) by expanding $|\phi(q_a)\rangle D(\Omega)$ in terms of the
Wigner $D-$functions\footnote{The definition of ${\mathcal{D}}^I_{M-K} (\Omega)$ used here
is that of Bohr-Mottelson}:
\begin{eqnarray}
\qquad|\psi_{IM}\rangle & = & \int d\tau d\Omega\ \delta(q-\hat Q) \sqrt{{%
\frac{2I+1}{16\pi^2}}} \sum_K \biggl[ |\phi_K(q_a)\rangle {\mathcal{D}}%
^I_{MK}(\Omega)  \nonumber \\
& & + (-1)^I |\phi_{-K}(q_a) {\mathcal{D}}^I_{M-K} (\Omega)\biggr] g_0
(q_a-\bar q_a)\ \ ,  \label{eq:9.2-3}
\end{eqnarray}
where $|\phi_K\rangle$ is an eigenfunction of $\hat{J_3}$ (\textit{i.e.}, $%
\hat{J}_3|\phi_K\rangle = K|\phi_K\rangle$) and ${\mathcal{D}}^I_{MK}$ is an
eigenfunction to ${\hat{\mathcal{L}}}_3$ (\textit{i.e.}, ${\hat{\mathcal{L}}}_3 {\mathcal{D}}%
^I_{MK} = K{\mathcal{D}}^I_{MK})$, ${\hat{\vec{\mathcal{L}}}}\,^2$ and ${\hat{\mathcal{L}}}_z$.
It then follows  that
\begin{equation}
{\hat{\vec J}}\,^2 |\psi_{IM}\rangle = I(I+1)|\psi_{IM}\rangle\ \ , \qquad\qquad J_Z
|\psi_{IM}\rangle = M|\psi_{IM}\rangle\ \ .  \label{eq:9.2-4}
\end{equation}
For the simpler case of the $K=0$ band, the wave function $|\psi_{IM}\rangle$ is given by:
\begin{equation}
|\psi_{IM} \rangle = \int d\tau d\Omega\ \delta(q- \hat Q) \sqrt{{\frac{2I+1}{%
8\pi^2}}} |\phi_0(q_a)\rangle {\mathcal{D}}^I_{M0}(\Omega) g_0(q-\bar q_a)\ \ .
\label{eq:9.2-5}
\end{equation}
Note that (as a consequence of axial symmetry) the following important
relation holds for both forms, (\ref{eq:9.2-3}) and (\ref{eq:9.2-5}), of $%
|\phi\rangle D(\Omega)$:
\begin{equation}
(\hat J_3 - {\hat{\mathcal{L}}}_3) |\phi\rangle D(\Omega) = 0 \ \ .  \label{eq:9.2-6}
\end{equation}

In this case of axial symmetry, and after omitting $%
E^0_{\mathrm{osc}} + E_{\mathrm{coupl}}$, we can see that the mean-energy (\ref{eq:9.2-2})  reduces to:
\begin{eqnarray}
\langle\psi_{IM}|\hat{H}|\psi_{IM}\rangle & = & \int d\tau d\Omega g^*_0(q_a -
\bar q_a) D^*(\Omega) \langle \phi (\bar q_a)|\delta(q-\hat Q)  \nonumber \\
& & \times \left[\hat{H}_0 +{\frac{B}{2}} \sum^2_{a=1} {\hat{\mathcal{L}}}^2_a\right]
|\phi(\bar q_a)\rangle g_0 D(\Omega)  \nonumber \\
& & + {\mathcal{E}}_1 + {\mathcal{E}}_2\ \ ,  \label{eq:9.2-7}
\end{eqnarray}
with
\begin{equation}
\hat{H}_0 = \hat{H} - \sum^2_{a=1} \dot{\hat Y} \hat J_a + {\frac{1}{2}} B\sum^2_{a=1}
\hat J^2_a\ \ ,  \label{eq:9.2-8}
\end{equation}
\begin{eqnarray}
{\mathcal{E}}_1 & = & \int d\tau d\Omega g^*_0D^*\langle\phi|\delta(q-\hat
Q)\sum^2_{a=1} (\dot{\hat Y}_a - B\hat J_a) {\hat{\mathcal{L}}}_a |\phi\rangle g_0
(q_a - \bar q_a) D(\Omega)\ \ ,  \nonumber \\
{\mathcal{E}}_2 & = & (B_{12} + B_{21}) \int d\tau d\Omega g^*_0
D^*\langle\phi|\delta(q- \hat Q) (\hat J_1 - {\hat{\mathcal{L}}}_1) (\hat J_2 -
{\hat{\mathcal{L}}}_2)|\phi\rangle g_0 D(\Omega) \ ,  \nonumber\\
\label{eq:9.2-10}
\end{eqnarray}
where $|\phi(q_a)\rangle D(\Omega)$ is given by (\ref{eq:9.2-3}) or (\ref%
{eq:9.2-5}), depending on whether one is interested in the $K\not=0$ band or
the $K=0$ band.

We should now specify the description of the intrinsic
structure of the system. To this end, we assume that the intrinsic state $%
|\phi(\bar q_a)\rangle$ is given by a mean field approximation such that $%
\langle\phi(\bar q_a)|\hat Q_{11} |\phi(\bar q_a) \rangle = \bar q_1$ is
equal to $\langle \phi|\hat Q_{22}|\phi\rangle = \bar q_2$ (\textit{i.e},
such that $\bar q_1 = \bar q_2$, the axial symmetry condition). This can be
achieved by means of a constrained variational principle.

Let us now look at the determination of the collective tensor operator $\hat Q$.
We determine the particle-hole $(ph)$
components of the tensor operator $\hat Q_{ab}$ such that $%
\langle\phi|\dot{\hat Y}_a - B\hat{J}_a|\phi\rangle$ is variationally stable,
\textit{i.e.},
\begin{equation}
\delta\langle\phi|\dot{\hat Y}_a - B\hat J_a |\phi\rangle = 0 \qquad
(a=1,2)\ \ .  \label{eq:9.2-11}
\end{equation}
This variational condition insures that the
simple expression (\ref{eq:9.2-3}) for $|\psi\rangle$ is adequate to
describe the rotational energy (term $\sim {\hat{\vec{\mathcal{L}}}}\,^2$) correctly.

To determine the mean energy (\ref{eq:9.2-7}), we need to calculate $%
{\mathcal{E}}_1$ and ${\mathcal{E}}_2$. In what follows, we are going to show
that both ${\mathcal{E}}_1$ and ${\mathcal{E}}_2$ are identically zero. First,
the term $\langle\phi|\delta(q- \hat Q)(\dot{\hat Y}_a - B\hat J_a)|\phi\rangle$
in the integrand of ${\mathcal{E}}_1$ can be rewritten as\footnote{%
In this approximate expression, we have neglected the two-body part of the
operator $\dot{\hat Y}_a\equiv \left[\hat H,i\hat Y_a\right]$.}
\begin{eqnarray}
\langle\phi|\delta(q- \hat Q)(\dot{\hat Y}_a - B\hat J_a) | \phi\rangle & \simeq &
\langle\phi|\delta(q- \hat Q) |\phi\rangle\langle\phi|\dot{\hat Y}_a - B\hat
J_a|\phi\rangle  \nonumber \\
& & + \sum_{\sigma\mu} \langle\phi|\delta(q- \hat Q)
|\phi_{\sigma\mu}\rangle\langle \phi_{\sigma\mu}|\dot{\hat Y}_a - B\hat J_a
|\phi\rangle\ \,\nonumber \\  \label{eq:9.2-12}
\end{eqnarray}
where $\sigma,\tau,\ldots$ refer to unoccupied (particle) states, while $%
\mu,\lambda,\ldots$ refer to occupied (hole) states. Using the condition (%
\ref{eq:9.2-11}), we see that the term $\langle\phi|\delta(q- \hat Q) (\dot{\hat Y}%
_a - B\hat{J}_a)|\phi\rangle$ becomes equal to $\langle\phi|\delta(q- \hat Q)
|\phi\rangle \langle\phi|\dot{\hat Y}_a - B\hat{J}_a|\phi\rangle$. Now, using the
fact that $|\phi_k\rangle$ is an eigenfunction to $\hat J_3$ and that the
action of both $\dot{\hat Y}_a$ and $\hat J_a$ $(a=1,2)$ on $|\phi_{k}\rangle$
generate $|\phi_{k\pm 1}\rangle$, we can ascertain that $\langle\phi|\dot{%
\hat Y}_a - B\hat J_a|\phi\rangle$ is itself identically zero, and hence $%
{\mathcal{E}}_1$ is equal to zero. To see this, note that $(\dot{\hat Y}_a -
B\hat J_a)$ have non-zero matrix elements only between $|\phi_k\rangle$ and $%
\langle\phi_{K\pm 1}|$. So,
\begin{equation}
\langle\phi_k|\dot{\hat Y}_a - B\hat J_a |\phi_k \rangle= 0\qquad (a=1,2)\ \
,  \label{eq:9.2-13}
\end{equation}
and also
\begin{equation}
\langle\phi_{-k}|\dot{\hat Y}_a - B\hat J_a |\phi_k\rangle = 0 \ \ ,
\label{eq:9.2-14}
\end{equation}
except for $K={\frac{1}{2}}$ but, in our case, $K$ is always an integer
(because we are dealing with an even nucleus).

Second, ${\mathcal{E}}_2$ is identically zero, since both $B_{12}$ and $B_{21}$
can be shown to be equal to zero. To see this, using these expressions,
\begin{eqnarray}
\qquad\qquad \left[\hat J_1,\ i\hat Y_1\right] = {\frac{\hat Q_{33} - \hat
Q_{22} }{q_3 - q_2}}\ \ ,& \qquad \left[\hat J_1,\ i\hat Y_2\right] = {\frac{%
\hat Q_{12} }{q_1 - q_3}}\ \ ,&   \label{eq:9.2-15} \\
\left[ \hat J_2,\ i\hat Y_2\right] = {\frac{\hat Q_{33} - \hat Q_{11} }{q_3 -
q_1}}\ \ ,& \qquad \left[\hat J_2,\ i\hat Y_1\right] = {\frac{-\hat Q_{12} }{%
q_2 - q_3}}\ \ , &  \label{eq:9.2-16}
\end{eqnarray}
we can easily show the following important relation:
\begin{equation}
\langle\phi|[\hat J_a,\ i\hat Y_b]|\phi\rangle = \delta_{ab}\qquad (a=1,2)\ \
,  \label{eq:9.2-17}
\end{equation}
since $\langle\hat Q_{ab}\rangle = \delta_{ab}$. Now, applying this relation
to the variational principle (\ref{eq:9.2-12}), we can verify that $%
\langle\phi|[\dot{\hat Y}_a,\ \hat Y_b] |\phi\rangle$ is equal to $B\delta_{ab}
$, \textit{i.e.},
\begin{equation}
B_{ab} = B\langle\phi|[\hat J_a,\  i\hat Y_b] |\phi\rangle = B\delta_{ab}\qquad
(a=1,2)\ \ .  \label{eq:9.2-18}
\end{equation}

Now, since both of  ${\mathcal{E}}_1$ and ${\mathcal{E}}_2$ are zero, and using the
relation $(\hat{J}_3 - {\hat{\mathcal{L}}_3)}|\phi\rangle D(\Omega) = 0 $ of (\ref%
{eq:9.2-6}) and (\ref{eq:9.2-11}), we can show that the mean energy (\ref%
{eq:9.2-7}) reduces to
\begin{eqnarray}
\langle\psi_{IM}|\hat{H}|\psi_{IM}\rangle & = & \int d\tau d\Omega
g^*_0(\Omega)\langle\phi|\delta(q-\hat Q) \left[\hat{H}-{\frac{1}{2}} B(\hat J^2_1
+ \hat J^2_2 + \hat J^2_3\right]|\phi\rangle g_0 D(\Omega)  \nonumber \\
& & + {\frac{B}{2}} \int d\tau d\Omega \langle\phi|\delta(q-\hat
Q)|\phi\rangle g^*_0 D^*(\Omega) \left[ {\hat{\mathcal{L}}}^2_1 +{\hat{\mathcal{L}}}^2_2 +%
{\hat{\mathcal{L}}}^2_3\right] g_0D(\Omega)\ .\nonumber \\   \label{eq:9.2-19}
\end{eqnarray}

Using the approximation
\begin{equation}
\langle\phi|\delta(q- \hat Q) \left( \hat{H}-{\frac{B}{2}}
{\hat{\vec J}}\,^2\right) |\phi\rangle\simeq \langle\phi|\delta(q- \hat Q)|\phi%
\rangle \langle\phi| \hat{H}-{\frac{B}{2}} {\hat{\vec J}}\,^2 |\phi\rangle,
\end{equation}
we can rewrite (\ref{eq:9.2-19}) in the following simpler form\footnote{%
Recall that we have omitted the vibrational energy part, $E^0_{\mathrm{osc}}$%
.}
\begin{eqnarray}
E_I = {\frac{\langle\psi_{IM}|\hat{H}|\psi_{IM}\rangle }{\langle
\psi_{IM}|\psi_{IM}\rangle}} \approx {\frac{1}{2}} BI(I+1) + \langle\phi|\hat{H} -
{\frac{1}{2}}B (\hat J^2_1 + \hat J^2_2 + \hat J^2_3)|\phi\rangle\ \ ,\nonumber \\
\label{eq:9.2-20}
\end{eqnarray}
where we have used he fact that ${\hat{\mathcal{L}}}^2{\mathcal{D}}^I_{MK} = I(I+1)%
{\mathcal{D}}^I_{MK}$. Note that the energy expression (\ref{eq:9.2-20}) has a
term, $-{\frac{1}{2}} B\langle{\hat{\vec J}}\,^2\rangle$ which represents a
substraction of a mean-rotational energy. This term is familiar from the
standard Peierls-Yoccoz angular momentum projection method. We expect this approximate treatment of the $\delta$-function in (\ref{eq:9.2-20}) to overestimate the mean-energy by a term of the order of half the zero-point vibration energy.

\vspace{0.5cm}

\noindent{\bf Moment of Inertia}\\
Let us now look at the moment of inertia, which is given by $B^{-1}$. The
inertial parameter $B$ can be determined from eqs. (\ref{eq:9.2-12}) and (%
\ref{eq:9.2-17}); \textit{i.e.}, it is given by the two equations
\begin{equation}
\langle\phi|\left[\hat H,\ i\hat Z_a\right] |\phi_{ph} \rangle =
\langle\phi|\hat{J}_a|\phi_{ph}\rangle\ \ ,\qquad\qquad B^{-1} = \langle\phi|\left[%
\hat Z_a,\ i\hat{J}_a\right]|\phi\rangle\ \ ,  \label{eq:9.2-21}
\end{equation}
where $\hat Z_a = \hat Y_a/B$. This expression for $B^{-1}$ is of the
well-known Thouless-Valantin form\cite{tval}. Note that if we neglect the
residual two-body interactions from the Hamiltonian, expressions (\ref%
{eq:9.2-21}) would give rise to Inglis cranking formula\cite{ingl}
\begin{equation}
{\frac{1}{B}} \approx 2\sum_{\sigma\mu} {\frac{|\langle\phi| \hat{J}_a
|\phi_{\sigma\mu}\rangle|^2 }{{\mathcal{E}}_\sigma - {\mathcal{E}}_\mu}}\ \ .
\label{eq:9.2-22}
\end{equation}
This approximate formula is well-known to overestimate the moment of inertia
quite badly.

In what follows, we are going to examine the connection between the BO
method and the large body of (rotational) data\cite{voig}, \textit{e.g.},
the moment of interia increases with angular momentum $I$. First, note that
Eq.~(\ref{eq:9.2-20}), which was derived for time-reversal
invariant $|\phi\rangle$, describe a rotational spectrum with constant
moment of inertia, $B^{-1}$, in disagreement with data. To see this,
consider the case $K=0$ for which $\langle\phi_0|\hat{J}_i|\phi_0\rangle = 0$ $%
(i=1,2,3)$. Hence, the energy expression (9.2.20) becomes
\begin{equation}
E_I = {\frac{1}{2}} BI(I+1) + \langle \hat{H}\rangle - \Delta E_{f\ell}\ \ ,
\label{eq:9.2-23}
\end{equation}
with $\Delta E_{f\ell} = {\frac{B}{2}} \langle \Delta \hat{J}^2_1 + \Delta
\hat{J}^2_2\rangle$, the fluctuation energy which is generated by angular momentum
fluctuations. In this case, therefore, the energy spectrum is that of a
rigid rotor, since the moment of inertia $B^{-1}$, as given by (\ref%
{eq:9.2-21}), is constant. This contradicts, of course, the experimental
facts.

Second, note that the failure of (\ref{eq:9.2-20}) to generate a moment of
inertia, $B^{-1}$, which increases with angular momentum is due to a
restrictive assumption on $|\phi\rangle$, the time-reversal invariance of $|\phi\rangle$ for $K=0$. In what follows, we are going to show that the BO approach has a
natural mechanism for introducing a moment of inertia which increases with $I
$, provided the restrictive assumption on $|\phi\rangle$ is dropped. In
addition, we will show that the energy we obtain for this case is lower than
the energy, (\ref{eq:9.2-23}), obtained with a time-reversal invariant $%
|\phi\rangle$. To this end, let us consider a symmetry-violating $%
|\phi\rangle$ for which $\langle \hat{J}_1\rangle$ is not zero but for which $%
\langle \hat{J}_2\rangle$ and $\langle \hat{J}_3\rangle$ are both zero. In this
analysis, we will restrict ourselves to the simplest case: $K=0$, and hence $%
|\phi_0\rangle$ is an eigenfunction to $\hat J_3$ with eigenvalue zero,
$\hat{J}_3|\phi_0\rangle \equiv 0$. In this case, the energy expression
(\ref{eq:9.2-20}) reduces to
\begin{equation}
E_I = {\frac{1}{2}} BI(I+1) + \langle \hat{H}\rangle - {\frac{B}{2}} \langle
\hat{J_1}\rangle^2 - \Delta E_{f\ell}\ \ .  \label{eq:9.2-24}
\end{equation}
Note that this energy is lower than the energy, (\ref{eq:9.2-23}), obtained
with a time-reversal invariant $|\phi\rangle$ (provided the fluctuation energy
is unchanged). In what follows, we shall neglect the angular momentum
fluctuations, $\Delta E_{f\ell}$, in the determination of the mean field $%
|\phi\rangle$. The energy expression (\ref{eq:9.2-24}) then provides a basis
for a variational determination of the symmetry violating $|\phi\rangle$:
\begin{equation}
\delta\langle \hat{H}- \omega  \hat{J_1}\rangle_\omega = 0 \ \ ,\qquad \omega = B\langle
\hat{J_1}\rangle_\omega\ \ ,  \label{eq:9.2-25}
\end{equation}
where the notation $\langle \hat 0\rangle_\omega$ is used to abbreviate $%
\langle\phi(\omega)|\hat 0|\phi(\omega)\rangle$. The parameter $\omega$ has,
obviously, the significance of an angular velocity.

\begin{figure}[tbp]
\centering
\thicklines
\begin{picture}(300,160)(-30,-10)
\put(0,0){\vector(1,0){230}}
\put(235,0){$\omega$}
\put(0,0){\vector(0,1){150}}
\put(-10,155){$\langle \hat{J_1}\rangle_\omega$}

\qbezier(0,0)(108,-3)(216,145)

\put(0,0.5){\line(2,1){220}}
\put(0,0){\line(2,1){220}}
\put(0,-0.5){\line(2,1){220}}
\qbezier(30,15)(40,5)(35,0)
\put(38,10){$\alpha$}

\put(83,0){\line(1,1){140}}
\qbezier(93,11)(100,10)(100,0)
\put(105,5){$\beta$}

\put(158,82){\line(1,0){15}}
\put(165,75){\line(0,1){15}}

\put(-5,82){\line(1,0){10}}
\put(-35,82){$\langle \hat{J_1}\rangle_\omega$}
\put(165,-5){\line(0,1){10}}
\put(160,-15){$ \omega$}

%
\end{picture}
\caption{$\langle \hat{J_1}\rangle_\omega$  increases with the angular
velocity $\omega$.}
\label{fig:rot1}
\end{figure}
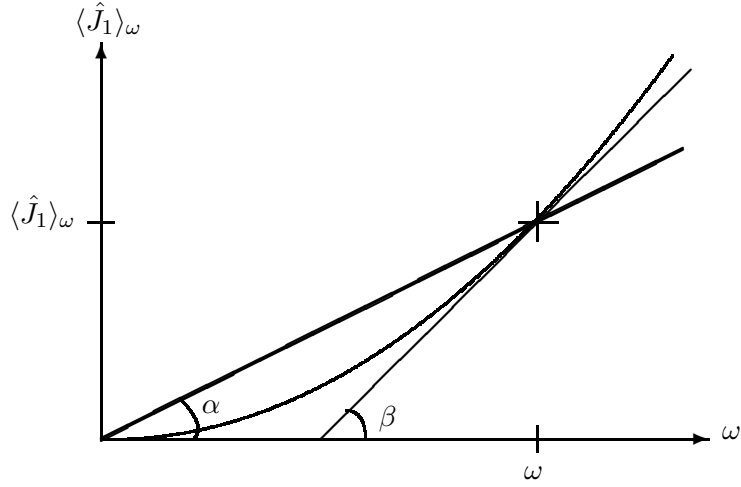

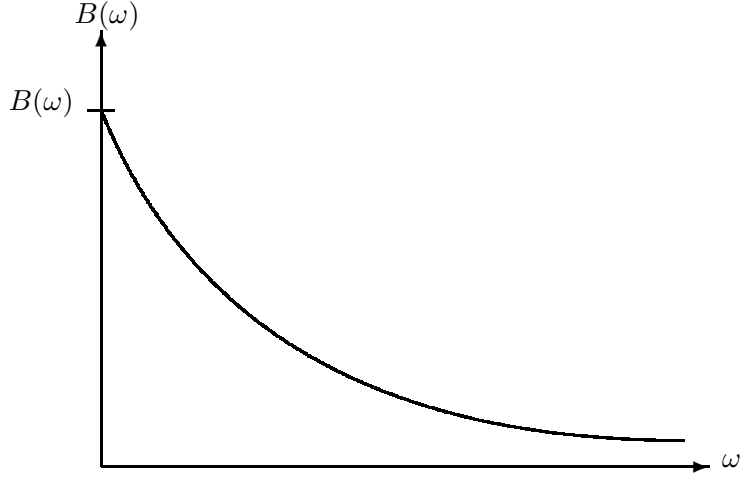
\begin{figure}[h]
\centering
\thicklines
\begin{picture}(300,165)(-30,0)
\put(0,0){\vector(1,0){230}}
\put(235,0){$\omega$}
\put(0,0){\vector(0,1){165}}
\put(-10,168){$B(\omega)$}

\put(-5,135){\line(1,0){10}}
\put(-35,135){$B(\omega)$}

\qbezier(0,135)(50,10)(220,10)

\end{picture}
\caption{The inertial parameter $B(\omega)$  decreases as the angular
velocity $\omega$ increases, since $dB/d\omega <0$. }
\label{fig:rot2}
\end{figure}

Now, we are in a position to show that $B(\omega)$ decreases when the
angular momentum increases. To see this, using the relation $B(\omega) = {%
\frac{\omega}{\langle \hat{J_1}\rangle_\omega}}$ of (\ref{eq:9.2-25}), we have
\begin{equation}
{\frac{dB(\omega) }{d\omega}} = {\frac{1}{\langle \hat{J_1}\rangle_\omega}} \left[
1 - {\frac{1}{\langle \hat{J_1}\rangle/\omega}}\ {\frac{d\langle \hat{J_1}\rangle_\omega%
}{d\omega}}\right]\ \ .  \label{eq:9.2-255}
\end{equation}
Since $\langle \hat{J_1}\rangle_\omega$ is well-known to increase with the angular
velocity $\omega$, and as shown in Fig. \ref{fig:rot1}, the slope ${\frac{d\langle \hat{J_1}\rangle_\omega}{d\omega}}$ is always larger than ${\frac{\langle \hat{J_1}\rangle_\omega }{\omega}}$. Thus, the slope, ${\frac{dB}{d\omega}}$ of $B(\omega)$ is
negative and, hence, $B(\omega)$ would behave as shown in Fig. \ref{fig:rot2}: $B(\omega)$  decreases as the angular velocity $\omega$ increases.

\noindent Therefore, we conclude that the moment of inertia $B^{-1}(\omega)$
increases, indeed, with angular velocity $\omega$, and hence with angular
momentum also.

In what follows, we are going to show that there exists a non-zero value, $%
\omega_c$, of $\omega$ at which the energy $E_I(\omega_c)$ of (\ref%
{eq:9.2-24}) is equal to its lowest value. To this end, let us write the
energy expression (\ref{eq:9.2-24}) in the following form (from which we
omit the fluctuation term, $\Delta E_{f\ell}$):
\begin{eqnarray}
\qquad\qquad E_I(\omega) &\simeq &\langle \hat{H}-\omega \hat{J_1}\rangle_\omega +{\frac{%
1}{2}} B(\omega) \left[ \langle \hat{J_1}\rangle^2_\omega + I(I+1)\right]
\nonumber \\
&=& \langle \hat{H}-\omega \hat{J_1}\rangle_\omega + {\frac{1}{2}} \left[ {\frac{\omega^2%
}{B(\omega)}} + B(\omega) I(I+1)\right]\ \ .  \label{eq:9.2-26}
\end{eqnarray}
First, note that the derivative,
\begin{equation}
{\frac{dE_I(\omega)}{d\omega}} = {\frac{1}{2}} \left[ I(I+1) - \langle
\hat{J_1}\rangle^2_\omega\right] {\frac{dB(\omega) }{d\omega}}\ \ ,
\label{eq:9.2-27}
\end{equation}
of $E_I(\omega)$ vanishes at a value $\omega_c$ which is determined by $%
\langle \hat{J_1}\rangle^2_{\omega_c} = I(I+1)$, \textit{i.e.},
\begin{equation}
{\frac{dE_I(\omega) }{d\omega}}\bigg|_{\omega_c} = 0\qquad \Longrightarrow \qquad \langle
\hat{J_1}\rangle^2_{\omega_c} = I(I+1)\ \ .  \label{eq:9.2-28}
\end{equation}
Second, we can easily show that the second derivative of $E_I(\omega)$,
\begin{equation}
{\frac{d^2E_I(\omega) }{d\omega^2}}\bigg|_{\omega=\omega_c} =\left. {\frac{\omega}{%
B^2(\omega)}} \left[ {\frac{\omega}{B(\omega)}} \left( {\frac{dB(\omega) }{%
d\omega}}\right)^2 - {\frac{dB(\omega)}{d\omega}}\right]\right|_{\omega=\omega_c}\ \
,  \label{eq:9.2-29}
\end{equation}
 is positive, since, as shown above (\textit{c.f.} Eq.~(\ref{eq:9.2-255})), ${\frac{dB}{d\omega}}$ is negative. Finally, we conclude
that, using a trial function $|\phi(\omega)\rangle$ whose time-reversal
symmetry is broken, one obtains, indeed, lower values for the inverse moment
of inertia, $B(\omega)$, and for the energy than those calculated with a $%
T\cdot R$ invariant mean field.

\vspace{0.1cm}

\begin{figure}[tbp]
\centering
\thicklines
\begin{picture}(300,235)(-30,-10)
\put(0,0){\vector(1,0){250}}
\put(255,0){$\omega$}
\put(0,0){\vector(0,1){215}}
\put(-25,220){$E_I(\omega)+\Delta E_{fl}$}

\qbezier(10,0)(100,-5)(250,115)
\put(-30,0){$E_0(0)$}

%
\qbezier(0,25)(2,25)(11,15)
\qbezier(11,15)(54,-17)(250,132)
\put(-30,25){$E_2(0)$}
\put(18,25){\line(1,0){15}}
\put(18,9){\line(1,0){15}}

\put(26,36){\vector(0,-1){11}}
\put(26,2){\vector(0,1){6}}

\put(18,14){$\Delta E_2$}

%
\qbezier(0,70)(5,70)(15,60)
\qbezier(15,60)(70,-10)(250,160)
\put(-30,70){$E_4(0)$}
\put(-40,39){$E_4(\omega_c)$}
\put(-5,39){\line(1,0){10}}
\put(44,39){\line(1,0){15}}
\put(44,70){\line(1,0){15}}
\put(52,49){\vector(0,-1){10}}
\put(52,60){\vector(0,1){10}}
\put(48,51){$\Delta E_4$}

%
\qbezier(0,180)(11,180)(25,150)
\qbezier(25,150)(70,60)(250,210)
\put(-40,115){$E_6(\omega_c)$}
\put(-5,115){\line(1,0){10}}
\put(-30,180){$E_6(0)$}
\put(67,115){\line(1,0){15}}
\put(67,180){\line(1,0){15}}
\put(75,135){\vector(0,-1){20}}
\put(75,160){\vector(0,1){20}}
\put(67,145){$\Delta E_6$}


\end{picture}
\caption{Behavior of $E_I(\omega)$  as a function of  the angular
velocity $\omega$, where $|\Delta E_I| = |E_I(0)-E_I(\omega_c)|$ is an increasing function of the angular momentum $I$}
\label{fig:rot3}
\end{figure}
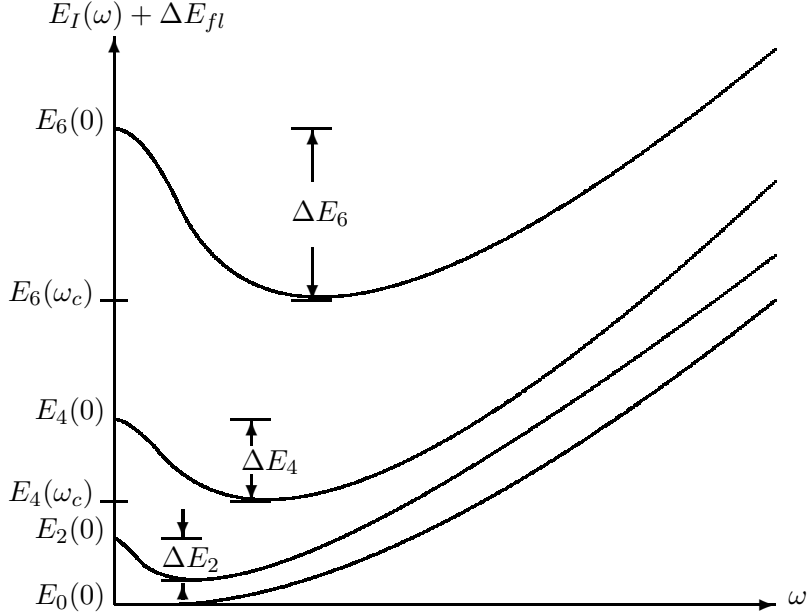

\noindent{\bf Calculation of the energy difference:}\ $\Delta E_I$\\
Let us now calculate the energy difference, $\Delta E_I$%
, between $E_I(\omega=0)$ and $E_I(\omega_c)$. Using Eq.~(\ref{eq:9.2-27}),
we can show that
\begin{eqnarray}
\Delta E_I & = & E_I(\omega_c) - E_I(0) = \int^{\omega_c}_0 {\frac{%
dE_I(\omega) }{d\omega}} d\omega  \nonumber \\
& = & {\frac{1}{2}} \left[ B(\omega_c) - B(0)\right] I(I+1) + {\frac{1}{2}}
\int^{\omega_c}_0 \omega^2 {\frac{d}{d\omega}} \left( {\frac{1}{B(\omega)}}%
\right) d\omega\ .\nonumber \\
\label{eq:9.2.30}
\end{eqnarray}
This expression can, after a partial integration, be reduced to
\begin{eqnarray}
 \Delta E_I & = & - {\frac{1}{2}} B_0 I(I+1) + \int^{\sqrt{I(I+1)}}_0
\omega\left(\langle \hat{J_1}\rangle\right) d\langle \hat{J_1}\rangle  \nonumber \\
& = & - \int^{\sqrt{I(I+1)}}_0\omega\left(\langle \hat{J_1}\rangle\right) \left[ {%
\frac{B}{B(\omega)}}-1\right] d\langle \hat{J_1}\rangle \ \ .  \label{eq:9.2-31}
\end{eqnarray}
This expression shows that $E_I(\omega_c)$ is, indeed, lower than $E_I(0)$,
since $B_0$ is larger than $B(\omega)$. So, if we know the dependence of the
angular velocity $\omega$ on $\langle \hat{J_1}\rangle_\omega$, we can easily
calculate the energy difference between $E_I(0)$ and $E_I(\omega_c)$. Note
that, the energy difference $|\Delta E_I|$ is an increasing function of the
angular momentum $I$. The qualitative behavior of the energy $E_I(\omega)$,
for various values of $I$, is plotted in Fig. \ref{fig:rot3}.

\section{Discussion and Conclusions}

Let us summarize what we have achieved in this work. First, we
have shown that the moment of inertia generated by the NBO method is
identical to the Thouless-Valantin form. Second, the two relations (\ref%
{eq:9.2-25}) and (\ref{eq:9.2-28}) determine the intrinsic
(symmetry-breaking) function $|\phi_\omega\rangle$ and the value, $\omega_c$%
, of $\omega$ where $E_I(\omega_c)$ is equal to its lowest value,
respectively. These two relations provide a bridge (connection) between the
NBO method, which is a truly quantum mechanical description of collective motion, and the
semi-classical approaches based on the idea of self-consistent cranking
(SCC). Thus, we have established a connection between the NBO approach and
the large body of experimental data, \textit{since the two relations (\ref%
{eq:9.2-25}) and (\ref{eq:9.2-28}) are known to provide reasonable
descriptions of vast amounts of empirical data ranging from low-lying
rotational states to high angular momentum states.}\cite{egid}--\cite{barn}
So, the present (NBO) method appears to be well-equipped to describe low as
well as high lying rotational states. Additionally, we should mention that
work has been started to apply the NBO method to the description of the
backbending phenomenon which was first observed by Johnson and his
collaborators.\cite{john}

In summary, we have studied here the rotational spectrum of even,
axially-symmetric nuclei within the framework of the NBO method. We have made
use of trial functions in which the intrinsic structure is described within  a
mean-field approximation. We have shown that the NBO formalism gives back the
Thouless-Valantin moment of inertia. Then, we have established a connection
between the NBO method and the SCC model, which has been successful in
reproducing vast amounts of experimental data. Finally, we have shown that
the introduction of a non time-reversal invariant intrinsic function both
lowers the energy for a given $I$, and provides a moment of inertia that
increases with the angular momentum $I$.

\vspace*{0.8cm}
\noindent{\Large\bf Acknowledgments}

Supported in part  by the Alabama Commission on Higher Education.

\end{document}